\documentclass[%
12pt,
reprint,
amsfonts,amsmath,amssymb,
aps,
prb,
floatfix,
]{revtex4-1}

\usepackage{mathptmx}
\usepackage{color}
\usepackage[tight]{subfigure}
\usepackage{graphicx}
\usepackage{dcolumn}
\usepackage{bm}
\usepackage[colorlinks,linkcolor={blue},citecolor={blue},urlcolor={blue}]{hyperref}
\usepackage{marginnote}
\usepackage{physics}

\DeclareSymbolFont{epsilon}{OML}{cmm}{m}{it}
\DeclareMathSymbol{\epsilon}{\mathord}{epsilon}{"0F}

\newcommand{\ba}{\begin{array}}
\newcommand{\ea}{\end{array}}

\newcommand{\ra}{\rangle}
\newcommand{\la}{\langle}

\newcommand{\veps}{\varepsilon}
\newcommand{\up}{\uparrow}
\newcommand{\dn}{\downarrow}
\newcommand{\nnum}{\nonumber}

\renewcommand{\vec}[1]{\mathbf{#1}}
\newcommand{\kv}{\mathbf{k}}
\newcommand{\rv}{\mathbf{r}}

\begin{document}
\title{Convolutional restricted Boltzmann machine (CRBM) correlated variational wave function
for the Hubbard model on a square lattice: Mott metal-insulator transition}
\author{Karthik V.}\email{karthik.v16@iisertvm.ac.in} 
\author{Amal Medhi}\email{amedhi@iisertvm.ac.in} 
\affiliation{Indian Institute of Science Education and Research Thiruvananthapuram,
Kerala 695551, India}

\begin{abstract}
We use a convolutional restricted Boltzmann machine (CRBM) neural network to construct a 
variational wave function (WF) for the Hubbard model on a square lattice and study it using
the variational Monte Carlo (VMC) method. In the wave function, the CRBM acts as a correlation 
factor to a mean-field BCS state.  The number of variational parameters in the WF does not grow 
automatically with the lattice size and it is computationally much more efficient compared to 
other neural network based WFs. We find that in the intermediate to strong coupling regime of the 
model at half-filling, the wave function outperforms even the highly accurate long range 
backflow-Jastrow correlated wave function. Using the WF, we study the ground state of the 
half-filled model as a function of onsite Coulomb repulsion $U$. We consider two cases for the 
next-nearest-neighbor hopping parameter, e.g., $t'=0$ as well as a frustrated model case with $t'\neq 0$.
By examining several quantities, e.g., double occupancy, charge gap, momentum distribution,
and spin-spin correlations, we find that the weekly correlated phase in both cases is paramagnetic 
metallic (PM). As $U$ is increased, the system undergoes a first-order Mott transition to an 
insulating state at a critical $U_c$, the value of which depends upon $t'$. 
The Mott state in both cases is spin gapped with long range antiferromagnetic (AF) order. 
Remarkably, the AF order emerges spontaneously from the wave function which does not
have any explicitly broken symmetry in it. Apart from some quantitative
differences in the results for the two values of $t'$, we find some interesting 
qualitative differences in the way the Mott transition takes place in the two cases. 
\end{abstract} 


\maketitle

\section{Introduction}
The Hubbard Hamiltonian has been studied extensively over the years using a variety of
analytical and numerical methods as a paradigmatic model for various correlated electron
phenomena\cite{Imada_RevModPhys.70.1039,Gull_PhysRevX.5.041041}. 
The model on a two-dimensional lattice is particularly interesting because of its relevance
to the high-temperature superconductors\cite{XiaoGang_RevModPhys.78.17}. Theoretically, the
model is handled well both in the weak and strong coupling limits by different methods,
but not so in the most interesting intermediate coupling regime. Among various numerical
methods, the variational Monte Carlo (VMC) technique has been a very useful
over the years in the study of the ground state properties of the model and its various 
extensions\cite{Ceperley_VMC_PhysRevB.16.3081,TaharaImada_VMC_JPSJ.77.114701,SorellaVMC,
Paramekanti_PhysRevB.70.054504,YokoyamaShiba_VMC1987}. Unlike other methods, VMC does not
suffer from any particular difficulty at any coupling strength, but its results are 
always biased by the choice of the variational wave function (WF) which is generally
constructed based on the ground state of an underlying mean-field Hamiltonian. 
However, with the advent of machine learning algorithms based on artificial 
neural networks (ANNs), 
it was realized that ANNs can also be used to represent a quantum many-body wave 
function \cite{Carrasquilla_NatPhys2017,
Carleo_Science2017,confusion,Broecker2017,Carrasquilla_PhysRevX.7.031038,
SDasDarma_PhysRevX.7.021021}. In such wave functions termed as neural-network quantum states (NQS),
the number of variational degrees of freedom is large and it offers the possibility to 
overcome the fundamental limitation of the variational method by enabling the construction of
a highly unbiased variational wave function.

A number of seminal works have already demonstrated the power of NQS to learn quantum 
many-body physics\cite{Carleo_RevModPhys.91.045002,Carrasquilla_PhysRevX.7.031038,
Carrasquilla_AdvPhys2020,Melko2019_NatPhys2019,Carrasquilla_PRXQuantum.2.040201}. To 
mention some of these, Carleo and Troyer\cite{Carleo_Science2017} first demonstrated 
the use of a restricted Boltzmann machine (RBM) based variational wave function to represent 
the ground state and study the dynamics of a few prototypical spin Hamiltonians accurately and 
efficiently. The method involved training the network by using a reinforcement learning mechanism 
which essentially is an iterative tuning of the network parameters so as to minimize
the variational energy. Torlai {\em et al.}\cite{Torlai_NatPhys2018} 
used an NQS wave function based on RBM network to perform a quantum state tomography 
and thereby learn the ground state of a quantum spin Hamiltonian. 
In Ref.~\onlinecite{Carleo_NatComm2018}, the authors used a wave function based on 
deep Boltzmann machine (DBM) and performed imaginary time evolution to obtain accurate 
ground state of the transverse-field Ising and the Heisenberg model.
Choo {\em et al.}\cite{Neupert_PRL2018}, by incorporating translational symmetries explicitly 
into the NQS wave function, managed to generate also the excited states for the Heisenberg 
and Bose-Hubbard model. RBM were also shown to be very efficient in representing 
topological quantum states due to the non-local geometry of its 
architecture\cite{Lu_PhysRevB.99.155136,Cirac_PhysRevX.8.011006}. 
However, most of these applications so far have been to bosonic systems. One main reason for this is that the 
fermionic systems have an additional complexity that comes from the non-trivial sign structures 
of its wave functions\cite{Weng_PhysRevB.90.165120}. The functions that can be represented by
neural-networks though highly non-linear, are essentially smooth and generally
fail to represent the sign structures of fermionic many-body wave functions. For instance, the
feed-forward neural networks, a type of ANN, were shown to be unable to correctly 
account for the sign structure of even a free fermionic wave function\cite{Cai_PhysRevB.97.035116}.
Nomura {\em et al.}\cite{Nomura_PhysRevB.96.205152} constructed a wave function 
by combining an RBM with a mean-field part where the RBM part introduces correlation effects,
a role usually played by Jastrow type correlation factors\cite{SorellaVMC}.
It was shown that the combination wave function give substantially lower energy than 
the conventional projected variational wave functions for the fermionic Hubbard model as well
as the Heisenberg model. There also exist other schemes based ANNs for fermionic systems
but these are restricted to small system size due to computational 
complexity\cite{Inui_PhysRevResearch.3.043126,Luo_PhysRevLett.122.226401,Pfau_PhysRevResearch.2.033429}.

In this work, we construct a ground state variational wave function for the fermionic Hubbard 
model on a square lattice using a convolutional restricted Boltzmann machine (CRBM) network where the 
CRBM is used as a correlation factor to a mean-field BCS state. We show that the wave function is 
not only computationally efficient compared to RBM wave function, it is also highly accurate in terms of the variational energy.  
Using VMC, we study the wave function as a ground state of the model at half-filling as a function of 
onsite Coulomb repulsion $U$. We consider two cases of the model when the next-nearest-neighbor 
hopping parameter $t'=0$ and when the model is frustrated with $t'\neq 0$. 
The wave function yields energies which are significantly
lower compared to those of the corresponding RBM or Jastrow projected wave functions, 
especially in the strong coupling limit. Indeed, in this limit it outperforms in terms of energy even
the highly accurate long range backflow-Jastrow correlated wave function for the 
model. The wave function correctly captures the presence of doublon-holon (DH) binding in the strong 
coupling regime in spite of it having no explicit DH binding term. It also give rise to long 
range antiferromagnetic (AF) order in the Mott insulating state spontaneously even though the wave 
function contains no explicit magnetic order. We examine the nature of Mott metal-insulator transition 
in the model by calculating various quantities, such as double occupancy, charge gap, 
momentum distribution and quasiparticle weight, and spin-spin correlation.
The results are thoroughly discussed and compared with other variational results.
The rest of the paper is organized as follows. In section~\ref{sec:model_method}, 
we describe the model and the CRBM wave function. In section~\ref{sec:results}, 
we describe and discuss the results, and finally in section~\ref{sec:conclusion}, 
we make the concluding remarks.

\section{Model and method}
\label{sec:model_method}
We consider the fermionic Hubbard model on a two-dimensional (2D) square lattice at half-filling.
The Hamiltonian is given by,
\begin{align}
\mathcal{H}=-\sum_{i,j\sigma}t_{ij}\left(c^\dag_{i\sigma}c_{j\sigma} + hc\right) 
+ U\sum_{i}n_{i\up}n_{i\dn}
\label{eq:hubbard_model}
\end{align}
where the operator $c^\dag_{i\sigma}$ creates an electron at site `$i$' with spin $\sigma$.
$c_{i\sigma}$ is the corresponding annihilation operator and $n_{i\sigma}=c^\dag_{i\sigma}c_{i\sigma}$ 
is the number operator. The first term represents hoppings of electrons from site to site.
The hopping integrals are $t_{ij}=t$ for $i$, $j$ nearest-neighbor (NN) sites, 
$t_{ij}=-t'$ for $i$, $j$ next-nearest-neighbor (NNN) sites, and zero otherwise.
Due to the particle-hole symmetry at half-filling, the model is equivalent to the case with 
opposite sign of $t'$. The Hamiltonian is the simplest paradigmatic model that 
capture the essential physics of several interesting correlated electron phenomenon in 
condensed matter physics, including Mott metal-insulator transition, high-temperature 
cuprate superconductors, etc.\cite{MottRevModPhys.40.677,Imada_RevModPhys.70.1039,
Orenstein468,XiaoGrangRevModPhys.78.17}. 
The Mott physics in the Hubbard model has been studied extensively using various methods 
\cite{DMFT_RevModPhys.68.13,Zhang_PhysRevLett.70.1666,
Rozenberg_PhysRevB.49.10181,Park_PhysRevLett.101.186403, Rozenberg_PhysRevLett.83.3498,
Bulla_PhysRevB.64.045103,Lu_PhysRevB.49.5687,Ferrero_PhysRevB.72.205126,Sigrist_EPJB2005,
SiddharthaLal_NJP2020}, including variational theory within the framework of various
Gutzwiller-Jastrow type wave functions (WFs)\cite{Yokoyama_VMC_III_JPSJ.59.3669,YokoyamaJPSJ2006,
Capello_PhysRevLett.94.026406,Capello_PhysRevB.73.245116,
Tocchio_PhysRevB.78.041101,Tocchio_PhysRevB.83.195138}. 
These projected variational wave functions are typically of the form, 
$|\Psi_{var}\rangle = {\cal P}\ket{\Phi}$, 
where $\ket{\Phi}$ is a one-body wave function which is generally taken to be ground state
of an underlying mean-field Hamiltonian.
${\cal P}$ is a projection operator or correlation factor which
introduces many-body correlation into the wave function. The simplest case is the 
Gutzwiller (GW) projector ${\cal P}_G$ which describes an on-site density-density correlation,
${\cal P}_G = \prod_i\left[1 - (1-g)n_{i\up}n_{i\dn}\right]$, $0 < g \le 1$.
It penalizes electronic configurations with doubly occupied sites thereby giving a better
description of the ground state of the Hubbard model compared to the uncorrelated state.
However, GW projector was found to be inadequate to describe the Mott insulating state.
This is because in the Mott state, local charge fluctuations creates configurations
with doubly occupied sites (doublon) next to empty sites (holon). The insulating nature of the 
state means that the doublons and holons must be bound to each other and
the GW projector does not capture this effect\cite{Kaplan_PhysRevLett.49.889,YokoyamaJPSJ2006,Capello_PhysRevLett.94.026406,Capello_PhysRevB.73.245116}. 
It was shown that the situation can be greatly improved by considering a correlation factor 
of the form ${\cal P} = {\cal P}_{DH}{\cal P}_G$, where the factor 
${\cal P}_{DH}$ incorporates doublon-holon (DH) binding in the wave function and  
is given by\cite{YokoyamaJPSJ2006},
${\cal P}_{DH} = \prod_{i}\left(1 - \eta Q_i\right)$, with
$Q_i = \prod_{\delta}\left[d_i(1-h_{i+\delta}) + h_i(1+d_{i+\delta})\right]$.
Here $d_i=n_{i\up}n_{i\dn}$ is doublon and $h_i=(1-n_{i\up})(1-n_{i\dn})$ is holon
operator, $\delta$ denotes the nearest-neighbor sites, and $\eta$ is a 
variational parameter. Alternatively, one can also consider Jastrow factor of 
the form\cite{Capello_PhysRevLett.94.026406},
${\cal P}_{J} = \exp\left(\sum_{ij}\frac{1}{2}v_{ij}(n_i-1)(n_j-1) +
w_{ij}h_id_j \right)$
which introduces long range correlations including DH binding. The description can be further 
improved by introducing a backflow correlation term in addition to the Jastrow factor and such 
a wave function considered for the half-filled Hubbard model was shown to be much better in
terms of the ground state energy and in the description of the Mott insulating 
state in the model\cite{Tocchio_PhysRevB.83.195138}.

Here, we consider a variational wave function where the correlator is 
based on an artificial neural-network (ANN). Such a wave function using the restricted 
Boltzmann machine (RBM) network as a correlator was already considered for the 
Hubbard as well as the Heisenberg model\cite{Nomura_PhysRevB.96.205152}. Here we consider a 
convolutional restricted Boltzmann machine (CRBM) network as a correlator. The CRBM is computationally
much more efficient than an RBM owing to much lesser number of network connections 
and consequently fewer number of variational parameters in CRBM. In fact, the number
of variational parameters in CRBM can be tuned and does not grow automatically with the
lattice size which is a much desirable feature computationally. In the followings, first
we describe the RBM network followed by a detailed description of the structure of the CRBM network.
The variational wave function using the CRBM as a correlator is described next.

\subsection{RBM network}
\begin{figure}[!htb]
\centering
\includegraphics[width=0.8\linewidth]{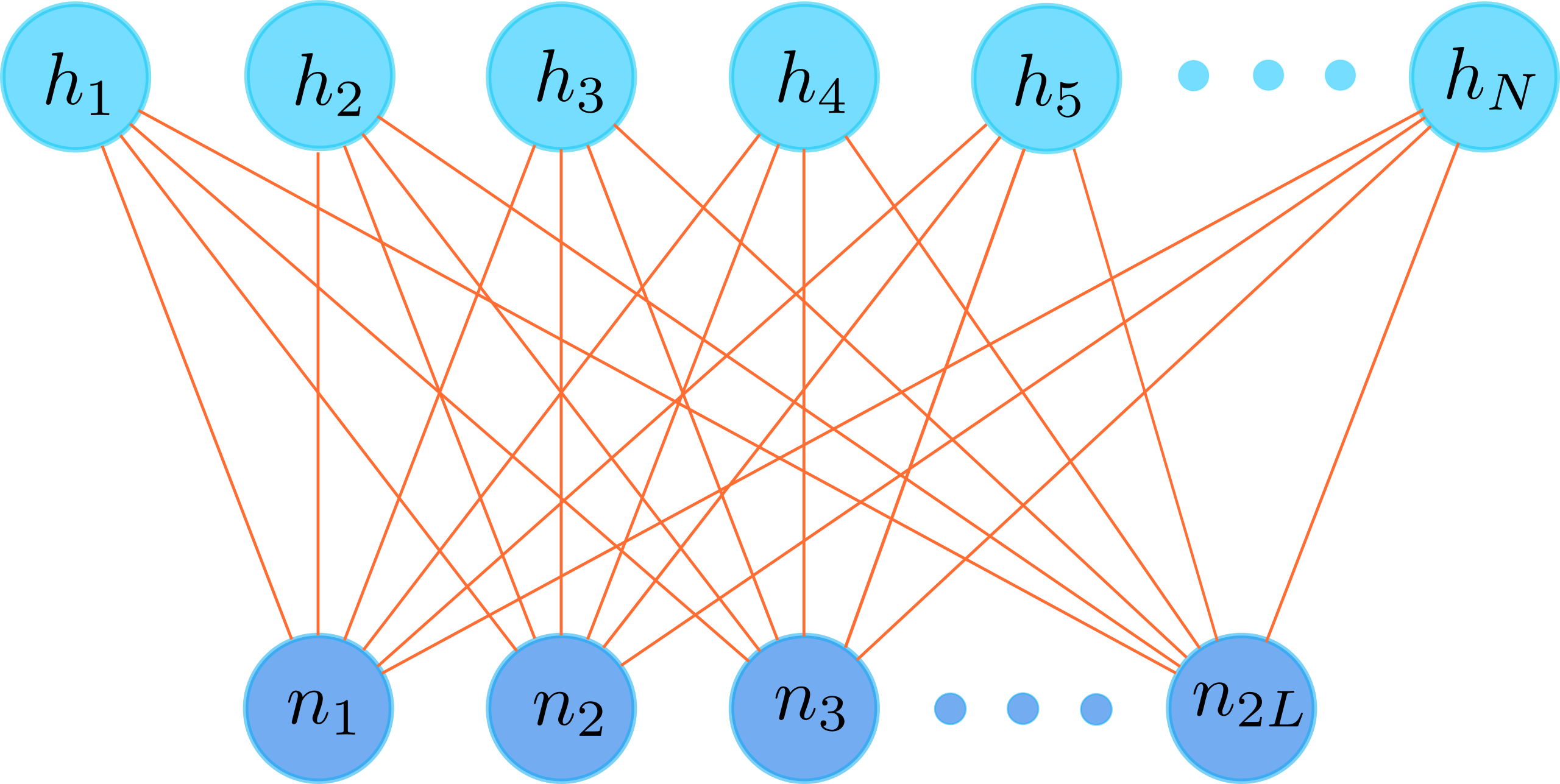}
\caption{A restricted Boltzmann machine (RBM) network. The artificial neurons in the 
visible layer take the inputs $\{n_1,n_2,\ldots,n_{2L}\}$ that represent the sets of 
electron occupation numbers. The neurons in the hidden layer define variables $h_i$
which can take values $\pm 1$.}
\label{fig:rbm}
\end{figure}

A restricted Boltzmann machine (RBM) network\cite{Carleo_NatComm2018} 
consists of two layers of artificial neurons, one visible layer connected to a hidden 
layer as shown in Fig.~\ref{fig:rbm}. There is no intralayer connection between the neurons. 
%
%
The number of neurons in the visible layer is $N_v=2L$ where $L$ is the number of lattice sites
and that in the hidden layer is $N_h$. 
The energy function of an RBM is given by,
\begin{align}
E_{RBM} =  -\sum_{i,j}w_{ij}h_in_j-\sum_ja_jn_j-\sum_ib_ih_i
\end{align}
The hidden variables $h_i$ take values $h_i=\pm 1$. The set $W=\{\bf a,b,w\}$ denotes
the set of all the network parameters $a_i$, $b_j$, and $w_{ij}$. Carrying out the summation
over the hidden variables, the probability distribution over the set of input 
values $\{n_1,n_2,\ldots n_{2L}\}$  is given by,
\begin{align}
\Psi_{RBM}(R,W)=\frac{1}{Z}\sum_{\bf h} e^{-E_{RBM}}
\end{align}
where $Z$ is the partition function. The number of parameters in the network can be 
drastically reduced by using the symmetries of the Hamiltonian. 
Here we take the hidden variable density to be $\alpha=N_h/N_v=1$. In this case, it can easily be
seen that imposition of lattice translational symmetry leads to a single bias parameter $a$ for
the visible units and a parameter $b$ for the hidden units. Also the number of unique 
elements in the weight matrix $w$ reduces to $2L$. Thus the total number of network 
parameters becomes $(2L+2)$ and the probability function reduces to,
\begin{align}
\Psi_{RBM}(R,{W})= \frac{1}{Z}e^{a\sum_in_i}\prod_{i} 2\cosh\Bigl(\sum_{j}w_{ij}n_j + b\Bigr)
\end{align}

\subsection{CRBM network}
\begin{figure}[!htb]
\centering
\includegraphics[width=0.8\linewidth]{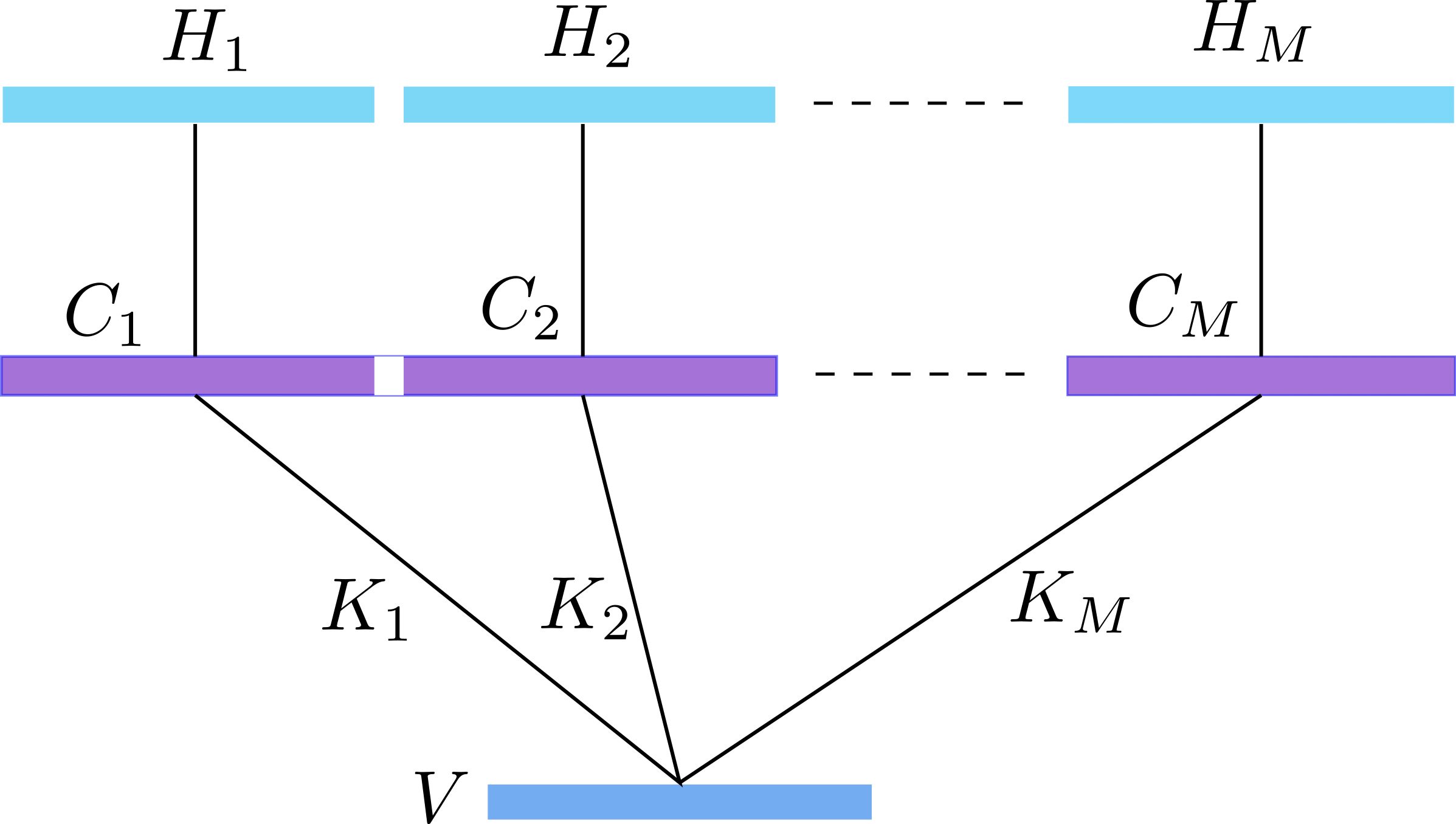}
\caption{Schematic diagram of a CRBM network. The visible layer ($V$) contains $N_v=2L$ 
neurons as in RBM.
The hidden layer is enlarged compared to RBM and consists of $M$ blocks $H_1,\,H_2,\,\ldots,H_M$, 
each containing $N_v$ neurons. There is a convolutional layer in between which also 
consists of $M$ blocks $C_1,\,C_2,\,\ldots,C_M$. The $m$-th block applies a convolutional filter 
with kernel $K_m$ to the visible layer and generates an output which is in the form of $N_v$ 
output neurons comprising the $C_m$ block. There are one-to-one connections between the neurons 
in the $C_m$ and $H_m$ blocks.}
\label{fig:crbm1}
\end{figure}

The convolutional restricted Boltzmann machine (CRBM) architecture is shown 
schematically in Fig.~\ref{fig:crbm1}. 
It can be thought of as consisting of 
three layers with an additional convolutional layer in between the visible and 
the hidden layers\cite{NorouziMastersthesis,Puente_PhysRevB.102.195148}. 
Since the underlying lattice is two-dimensional (2D), it is necessary
to consider layers also to be 2D arrays of neurons instead of the 
linear arrangement of Fig.~\ref{fig:rbm}.
The visible layer again contain $N_v=2L$ neurons, like in RBM.
But the hidden layer ($H$ layer) in CRBM is enlarged to make it into $M$ blocks 
$H_1,\,H_2\,\ldots,H_M$, each block containing $N_h=N_v$ neurons. 
The middle convolutional layer ($C$ layer) also consists of $M$ blocks $C_1,\,C_2,\,\ldots,\,C_M$. 
The input layer, the blocks $C_m$-s, $H_m$-s are all taken to be $2P\times P$ arrays of 
neurons, where $P=\sqrt{L}$.
The $m$-th block applies a convolution filter with kernel $K_m$ to the input layer
and produces a result which is in the form of $N_v$ number of convolutional output 
neurons ($2P\times P$ array) comprising the $C_m$ block. 
The convolutional output blocks are connected directly to the hidden layer blocks. That 
is, the $i$-th neuron in $m$-th convolutional block is connected only to the $i$-th neuron of
the $m$-th hidden layer block, and not to any other hidden unit. 
Thus, there is a one-to-one connection between the neurons in the blocks $C_m$ and $H_m$, 
with a total of $N_v$ connections between the two sub-layers. Therefore for $M\ll L$, the total
number of connections in the CRBM is much less compared to that in the RBM, 
making the CRBM representation much more efficient in comparison.
\begin{figure}[!htb]
\centering
\includegraphics[width=0.7\linewidth]{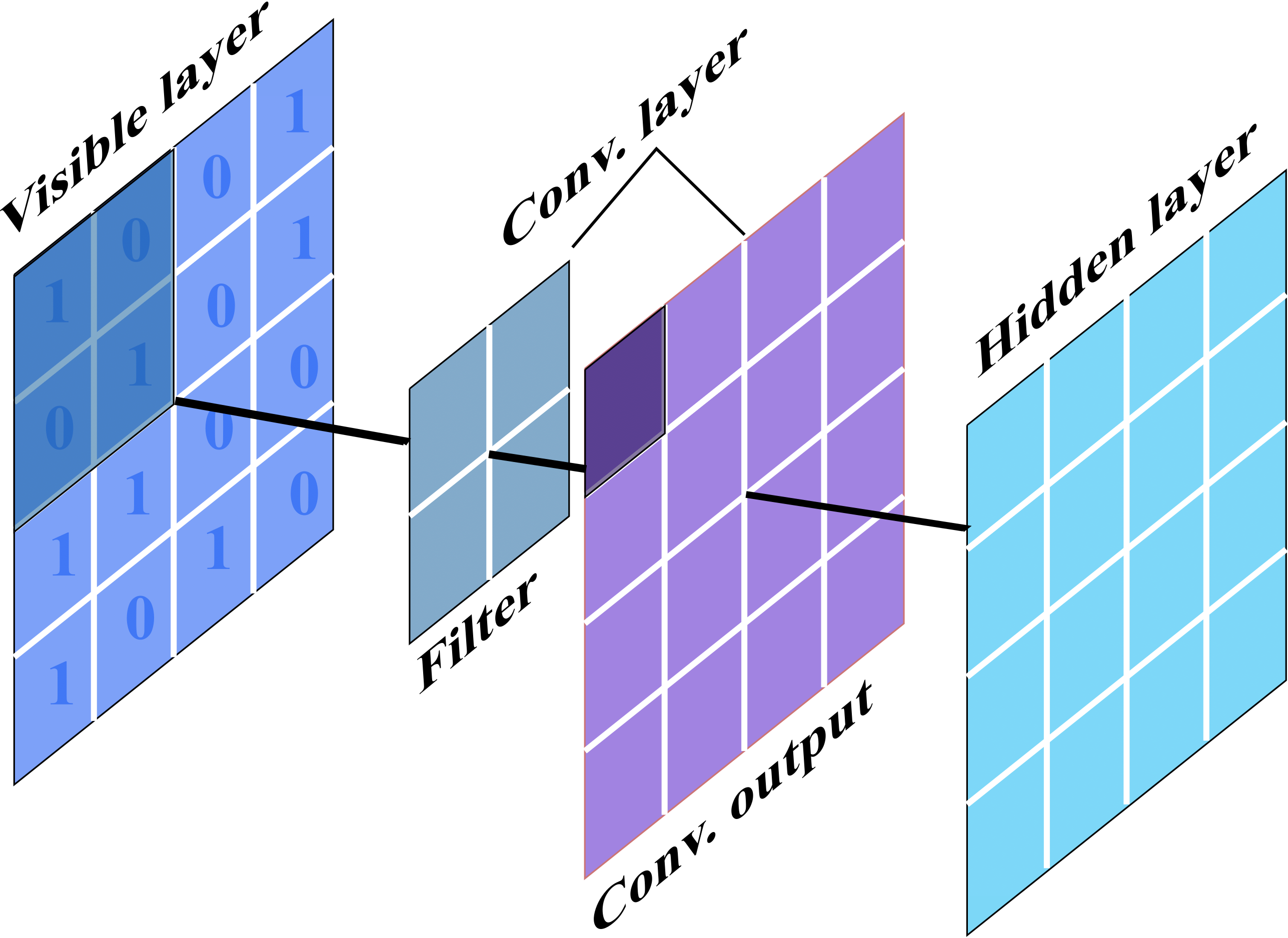}
\caption{Schematic diagram showing the convolutional operation by one of the $C$-blocks in 
Fig.~\ref{fig:crbm1}. The convolutional layer slides a filter over the visible 
layer with a stride of one. The operation generates the convolutional output layer
which is of the same dimension as the input layer.}
\label{fig:crbm2}
\end{figure}
Next, consider the convolutional operation by the $m$-th block (Fig.~\ref{fig:crbm2}). 
The convolutional filter in the block is defined by its kernel $K_m$ which is a
matrix of $D\times D$ parameters ($D\leq P$). Action of the filter on a $D\times D$ 
window of the visible layer is given by the generalized dot product $K_m\odot V_k$, where 
$V_k$ is the matrix formed by the input values given to the neurons in this window.
This is shown schematically in Fig.~\ref{fig:crbm2}. We slide the filter with a stride
of one step along both the directions to cover the whole of the visible layer.
This creates $N_v$ number
of windows of the visible layer and hence $N_v$ number of convolutional outputs 
$K_m\odot V_k$, $k=1,\,2,\,\ldots N_v$. Denoting the $i$-th hidden variable in the $m$-th 
block by $h_{mi}$, the joint energy function in the CRBM is given by,
\begin{align}
E_{CRBM} = -\sum_{m,k}h_{mk}(K_m\odot V_k) - a\sum_{j}n_{j} - \sum_{mi}b_mh_{mi}
\end{align} 
It may be mentioned that by construction CRBM conservs the lattice translational symmetry
imposed in the corresponding RBM. The probability distribution over the visible variables
in CRBM is given by,
\begin{align}
\Psi_{CRBM} =& \frac{1}{Z}\sum_{\bf h}e^{-E_{CRBM}} \nnum\\
=& \frac{1}{Z} e^{a\sum_j n_j} \prod_{mk} 2\cosh\left(K_m\odot V_k + b_m\right)
\label{eq:psi_crbm}
\end{align} 
where $Z$ is a normalization constant. The total number of parameters in the CRBM is
$M(2D^2+1)$. This number depends only upon the two network structural parameters $M$ and $D$, 
and not upon the lattice size, $L$. Therefore the number of variational parameters in the
CRBM wave function does not grow directly with the lattice size which is a big advantage 
in the optimization process. In practice, the parameters $M$ and $D$ are determined by 
tuning its values so as to obtain the best variational energy. The lattice size might affect 
these values somewhat, but the number of variational parameters is still expected to be much 
less than that in the corresponding RBM wave function.

\subsection{CRBM correlated wave function}
The CRBM correlated wave function that we consider here is given by,
\begin{align}
|\Psi_{var}\rangle = {\cal P}_{CRBM}\ket{BCS}_N
\end{align}
where $\ket{BCS}_N$ is the ground state (with fixed number of particles) of the 
following mean-field Hamiltonian,
\begin{align}
{\cal H}_{MF} =& \sum_{\kv\sigma}\veps_{\kv}c^\dag_{\kv\sigma}c_{\kv\sigma} 
 -\sum_{\kv}\left(\Delta_{\kv}c^{\dag}_{\kv\up}c^{\dag}_{-\kv\dn}+hc\right) 
\end{align}
Here $\veps_{\kv} = -2t(\cos k_x+\cos k_y)+4t'\cos k_x\cos k_y-\mu$, $\mu$ being the
chemical potential. We take superconducting pairing amplitude $\Delta_{\kv}$ to be of 
$d_{x^2-y^2}$-wave ($d$-wave) symmetry with $\Delta_{\kv} = \Delta_{SC}(\cos k_x-\cos k_y)$,
where $\Delta_{SC}$ is the SC gap parameter. 
The quantities $\Delta_{SC}$ and $\mu$ are the variational parameters in the 
one-body part of the wave function. 
We consider the wave function in fixed particle number ($2N$) representation with equal
number of up and down spins. 
In terms of real-space (Wannier) basis, $\ket{BCS}_N$ can be 
expressed as,
\begin{align}
\ket{BCS}_N =& \sum_{n_1,n_2,\ldots n_{2L}}  \Psi_{BCS}(n_1,n_2,\ldots n_{2L}) \nnum\\
& \times (c^{\dag}_{1\up})^{n_1}\ldots (c^{\dag}_{L\up})^{n_L}
 (c^{\dag}_{1\dn})^{n_{L+1}}\ldots (c^{\dag}_{L\dn})^{n_{2L}}\ket{0}  \nnum\\
 =& \sum_R\Psi_{BCS}(R)\ket{R} 
\end{align}
Here $n_i$-s are the occupation numbers which can take value $0$ or $1$. $L$ is the
number of lattice sites. The summation is over the set of values 
$\{n_1,\ldots n_{2L}\}\equiv R$ subject to constraint $\sum_{i=1}^L n_i=\sum_{i={L+1}}^{2L}n_i=N$.
The amplitudes $\Psi_{BCS}(R)$-s are the determinantal coefficients corresponding to the
electronic configurations. The action of ${\cal P}_{CRBM}$ is given by,
\begin{align}
\ket{\Psi_{var}} = {\cal P}_{CRBM}\ket{\Phi} = \sum_R \Psi_{CRBM}(R)\Psi_{BCS}(R)\ket{R}
\label{eq:wf_finalform}
\end{align}
where $\Psi_{CRBM}(R)$ is output (Eq.~(\ref{eq:psi_crbm})) of the CRBM network for a given set of 
input values $R=\{n_1,\ldots n_{2L}\}$. The variational parameters in the wave function
consists of the network parameters plus the parameters in the mean-field part of the 
wave function.

\subsection{Variational Monte Carlo}
Having constructed the variational wave function, we use the variational Monte Carlo (VMC) 
method\cite{Ceperley_VMC_PhysRevB.16.3081,SorellaVMC,TaharaImada_VMC_JPSJ.77.114701} to
compute the variational energy,
\begin{align}
E_{var}(\bm{\alpha}) = \frac{\bra{\Psi_{var}}{{\cal H}}\ket{\Psi_{var}}}{\bra{\Psi_{var}}\Psi_{var}\rangle}
\end{align}
and minimize it with respect to the variational parameters $\bm{\alpha}$. 
We use the stochastic reconfiguration (SR) technique\cite{SorellaVMC,
TaharaImada_VMC_JPSJ.77.114701} for optimization which generally works all even 
for large number of variational parameters. In this method, the variational 
parameters $\bm{\alpha}$ are updated as,
$\bm{\gamma} = \bm{\alpha} - \Delta t S^{-1} \vec{g}$
where $\vec{g}=\bm{\nabla}_\alpha E_{var}$ is the energy gradient. $S$ is the overlap 
matrix with the matrix elements given by, $S_{kl} = \langle \Psi_{\alpha_k} \ket{\Psi_{\alpha_l}}$,
$\ket{\Psi_{\alpha_k}} = \frac{\partial}{\partial \alpha_k}\ket{\Psi_{\bm{\alpha}}}$.
In the VMC simulations, typically we take $\sim 10^5$ measuring steps after warming up the system for 
$\sim 10^3$ steps. Each MC step consists of $L$ number of MC moves that
include both the hopping and exchange moves as mentioned before. The number of variational
parameters in the CRBM wave function depends upon the two network structural parameters 
$M$ and $D$, and independent of the lattice size. For the values of $M$ and $D$ 
considered here, the number of variational parameters becomes of the order of 150. The
optimization step even with the SR method sometime does need large number of iterations to
converge which creates a bottleneck in the computations.

\section{Results}
\label{sec:results}
We consider the Hamiltonian on a square lattice of size $10\times 10$ with $L=100$ sites,
and a band filling of one particle per site (half-filling). We consider two values of the 
next-nearest hopping parameter $t'$, 
e.g.\ $t'/t=0$ and $t'/t=0.5$. The non-zero value $t'/t$ brings in frustration in the 
antiferromagnetic order expected at half-filling. We study the model as a function of 
Hubbard onsite interaction $U/t$. In the results below, all the energy values quoted are 
energy per site and in units of $t$.

\begin{figure}[!htb]
\centering
\includegraphics[width=0.50\columnwidth]{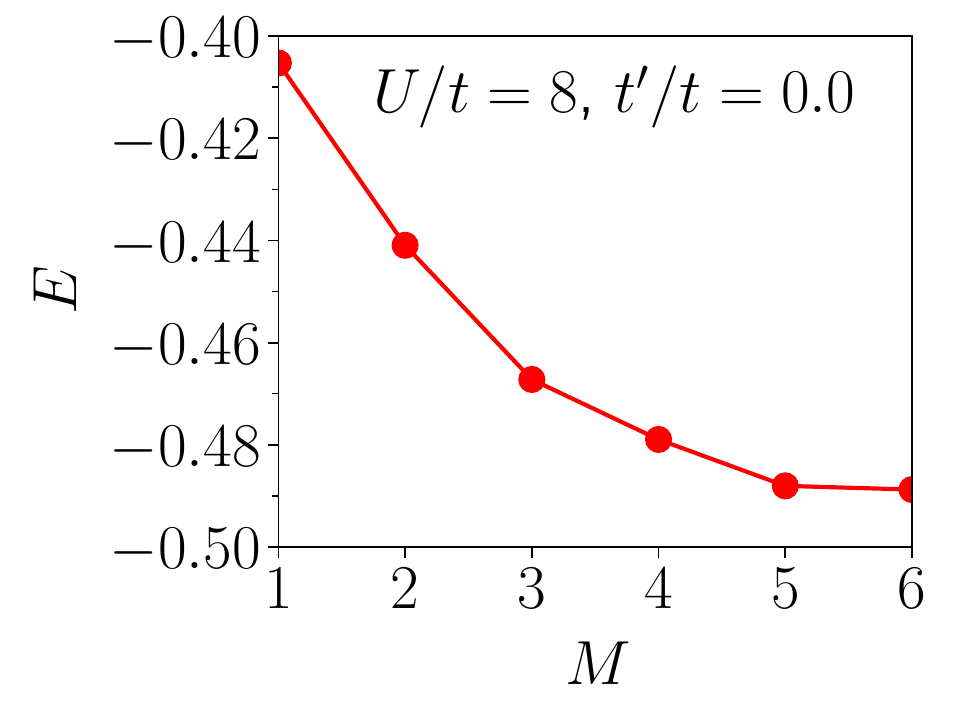}
\caption{Energy obtained by repeating optimizations for different values of the 
CRBM parameter $M$. The energy reaches a saturation value 
at $M\sim 5$. The value of the parameter $D$ is 4. Model parameters, $U/t=8$ and $t'/t=0$.}
\label{fig:e_vs_M}
\end{figure}

Before going ahead with the calculations, we need to determine the optimal 
configuration of the CRBM network. It has two crucial structural
parameters - the dimension $D$ of the convolutional kernel and the number of 
convolution blocks $M$. In principle, $D$ can vary from 1 to $L$. The convolution can be 
interpreted as a `feature extraction' operation\cite{NorouziMastersthesis}. A value $D=1$ 
implies a trivial operation while $D=L$ implies flattening of all features. Here we 
take $D=4$ which we find to be an optimal value in terms of performance and efficiency.
Regarding value of $M$, we checked the energy obtained by repeating the optimizations 
for different values of $M$. An example plot is shown in Fig.~\ref{fig:e_vs_M}. 
It shows that at $U/t=8$, the best energy is obtained for a minimum $M$ value of $5$.
In fact, the optimal value of $M$ depends upon the value of $U$. It is smaller for 
smaller $U$. Therefore, a value of $5$ works well for the range of $U/t$ considered
here and hence we set $M=5$ for the rest of the calculations.

\begin{figure}[!htb]
\centering
\subfigure[\label{fig:en_comp_a}]{
\includegraphics[width=0.48\columnwidth]{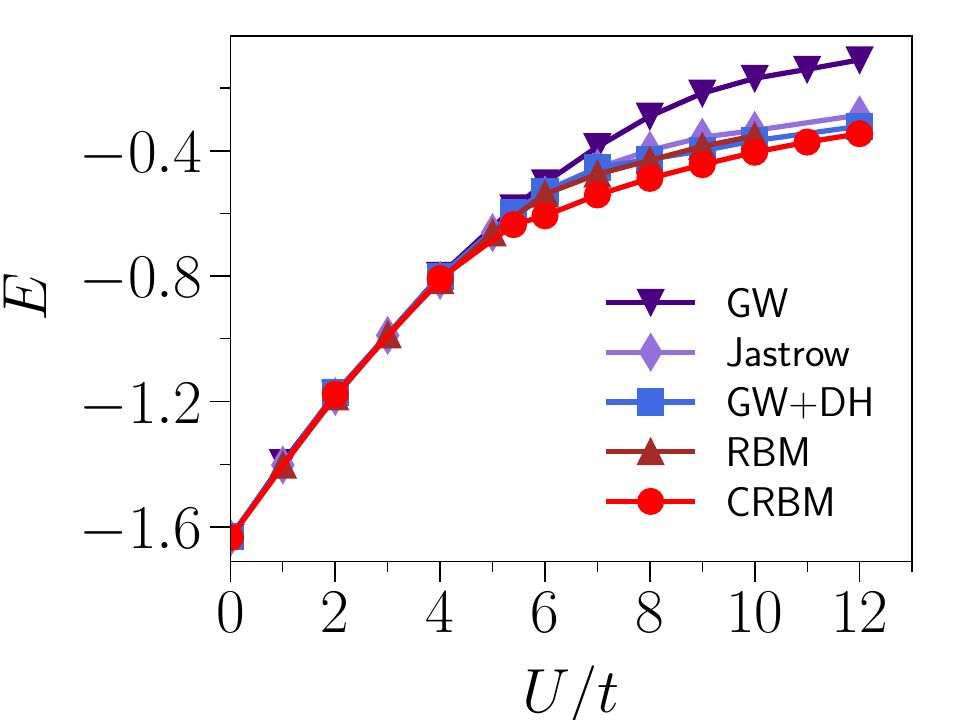}}
\subfigure[\label{fig:en_comp_b}]{
\includegraphics[width=0.48\columnwidth]{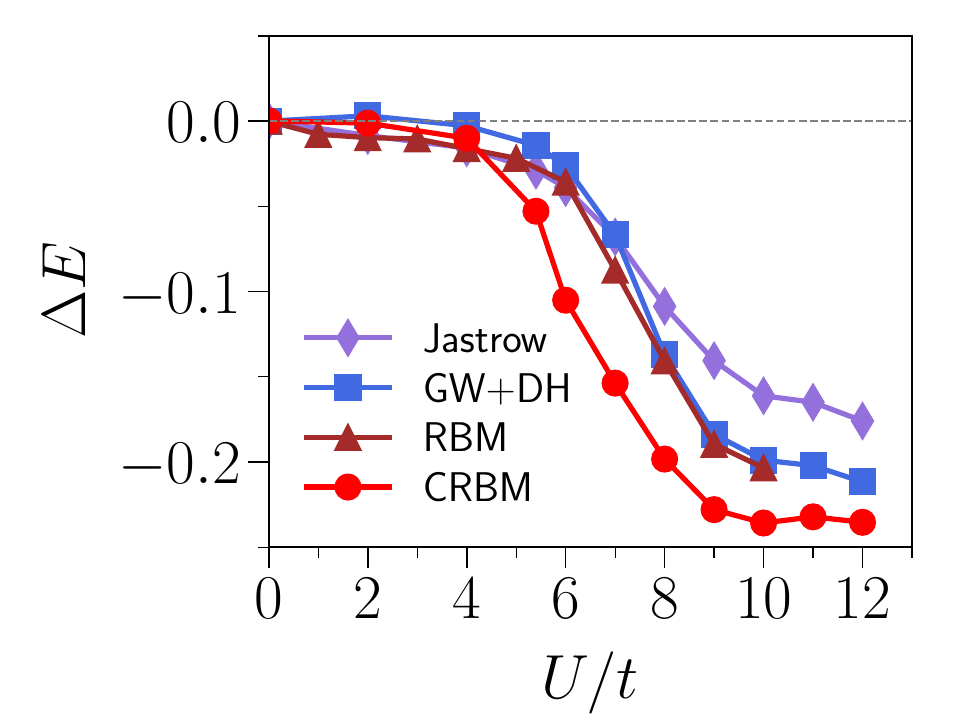}}
\caption{(a) Energies of the five wave functions described in the text as a 
function of $U/t$. (b) The difference $\Delta E = E_{X}-E_{GW}$, where $X$ stands
for the four wave function names shown in the figure.}
\label{fig:en_comp}
\end{figure}

We optimize the CRBM wave function defined in
Eq.~(\ref{eq:wf_finalform}) by minimizing the corresponding variational energy for a 
range of model parameter values. For comparison, we also calculate the energies
of four other variational wave functions. These are (i) $\ket{\Psi_{GW}}={\cal P}_G\ket{BCS}_N$ (GW)
(ii) $\ket{\Psi_{GW+DH}}={\cal P}_G{\cal P}_{DH}\ket{BCS}_N$ (GW+DH) 
(iii) $\ket{\Psi_{Jastrow}}={\cal P}_J\ket{BCS}_N$ (Jastrow) and 
(iv) $\ket{\Psi_{RBM}}={\cal P}_{RBM}\ket{BCS}_N$ (RBM), where the projection operators 
are as defined above. In $\ket{\Psi_{RBM}}$, we use an RBM network as the correlator.
The comparison of energies of these wave functions are shown in Fig.~\ref{fig:en_comp}.
It shows that, in the weak coupling limit, the energies of these WFs are more or less 
similar with minor differences among them. However, the energies start to differ in
the strong coupling regime. This is clear from Fig.~\ref{fig:en_comp_b} where we have plotted 
the difference $\Delta E=E_{X}-E_{GW}$, where $X$ stands for the four other wave 
functions shown in the figure. As the figure shows, $\Delta E$ is negative in the
intermediate to strong coupling regime, indicating that the energy of the GW wave function
is highest here. The energies of the GW+DH and RBM wave functions are comparable and lower than 
that of the Jastrow wave function. The best wave function among the five is the CRBM 
wave function which yields energies significantly lower than the other four. 
We also compare the CRBM energies with those for the long range backflow-Jastrow correlated wave
function used in Ref.~\onlinecite{Tocchio_PhysRevB.83.195138}, which was shown to be the most accurate
among the Jastrow type projected wave functions. We find that the CRBM wave function even 
outperforms the backflow-Jastrow wave function. For example, the backflow WF give 
energies per site equal to $-0.5961(1)$, $-0.4803(1)$, $-0.4022(1)$ and
$-0.3451(1)$ at $U/t$ equal to $6$, $8$, $10$, and $12$, respectively. For the same
values of $U/t$, the CRBM wave function give energies per site equal to
$-0.6076(2)$, $-0.4882(2)$, $-0.4048(2)$, and $-0.3456(2)$, respectively. These 
energies are clearly lower than the backflow WF energies. It must be mentioned that
the lattice sizes used in these two studies are not the same, and hence the 
comparison is not strictly rigorous. Still, it gives an idea about how accurate the
CRBM wave function is. In Ref.~\onlinecite{Nomura_PhysRevB.96.205152}, the authors used 
an RBM correlator in conjunction with a many-variable one-body wave function. The energies 
of this wave function is actually lower than the CRBM energies, though such a
wave function is computationally much more expensive.

Next, we examine the ground state phase of the model as a function of $U/t$ as 
described by the CRBM wave function. As mentioned before, we have done 
calculations for two different values of $t'$, e.g. $t'/t=0$ and $t'/t=0.5$. 
Although the model has been studied extensively in the past using a variety of methods, 
several important aspects of its phase diagram including the nature of the zero temperature Mott 
transition, are not yet established 
unambiguously\cite{Toschi_PhysRevB.91.125109,Gull_PhysRevX.5.041041,SiddharthaLal_NJP2020}.
For example, questions exist whether the ground state of the unfrustrated model ($t'/t=0$) 
on square lattice is antiferromagnetic (AF) insulating at any $U/t>0$, or is it 
paramagnetic with a finite value of critical interaction $U_c/t$ for Mott metal-insulator
transition. What is the nature of Mott insulating state, is it magnetically ordered
or a spin liquid?
If the unfrustrated model is AF insulating, is there a critical value of $t'$ 
beyond which the ground state become paramagnetic metallic? The answers to these questions 
somewhat vary among different methods. Within the variational theory, the phase diagram 
of the two-dimensional Hubbard model has been studied by looking at the competition
between wave functions with different symmetries. In these wave functions, 
the one-body part is taken to be either pure BCS type without any magnetic order or 
an AF type with explicitly broken spin rotational symmetry. By using the GW-DH 
projector as the correlator, Yokoyama {\it et al.}\cite{YokoyamaJPSJ2006} have 
shown that at $t'=0$, the lowest energy state at any $U/t>0$ is the symmetry broken one with 
long range AF order and insulating. For $t'/t>0$, there
exists a non-zero value of critical interaction $U_c$ below which the state
is paramagnetic metallic (PM). For $U/t>U_c/t$, the state is AF insulating for small $t'/t$,
but paramagnetic insulating at large $t'/t$. The transition to the AF insulating state
at $t'/t=0$ is continuous. It gradually turns first-order at larger $t'/t$. However, 
if considered within the non-magnetic projected BCS wave function only, 
the state is found to be PM for any $t'/t$ below a critical $U_c/t \neq 0$. In this case, 
the Mott transition is of first-order nature at any $t'/t$, and the insulating state is 
non-magnetic. In another work, Tocchio {\it et al}\cite{Tocchio_PhysRevB.78.041101} 
used variational WFs with backflow correlations in addition to a long range 
Jastrow projector. The backflow correlated WFs is much more accurate with 
lower variational energy compared to only DH or Jastrow projected WFs. The study
also showed that for small $U/t$ and non-zero $t'/t$,
the ground is paramagnetic metallic. It becomes insulating with a long range AF order
as $U$ is increased above a critical value. Interestingly, the backflow WF give
a region in the phase diagram at large enough values of $U/t$ and $t'/t$, where the state is 
insulating spin-liquid without any long range magnetic order.

\begin{figure}[!htb]
\centering
\subfigure[\label{fig:dh_corr_a}]{
\includegraphics[width=0.46\columnwidth]{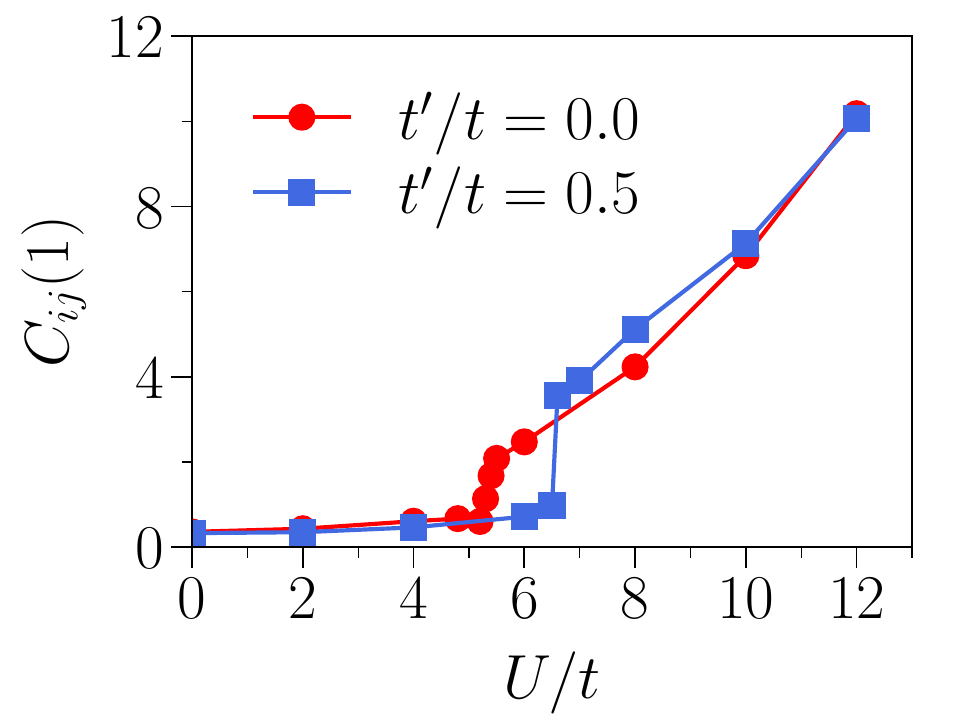}}
\subfigure[\label{fig:dh_corr_b}]{
\includegraphics[width=0.46\columnwidth]{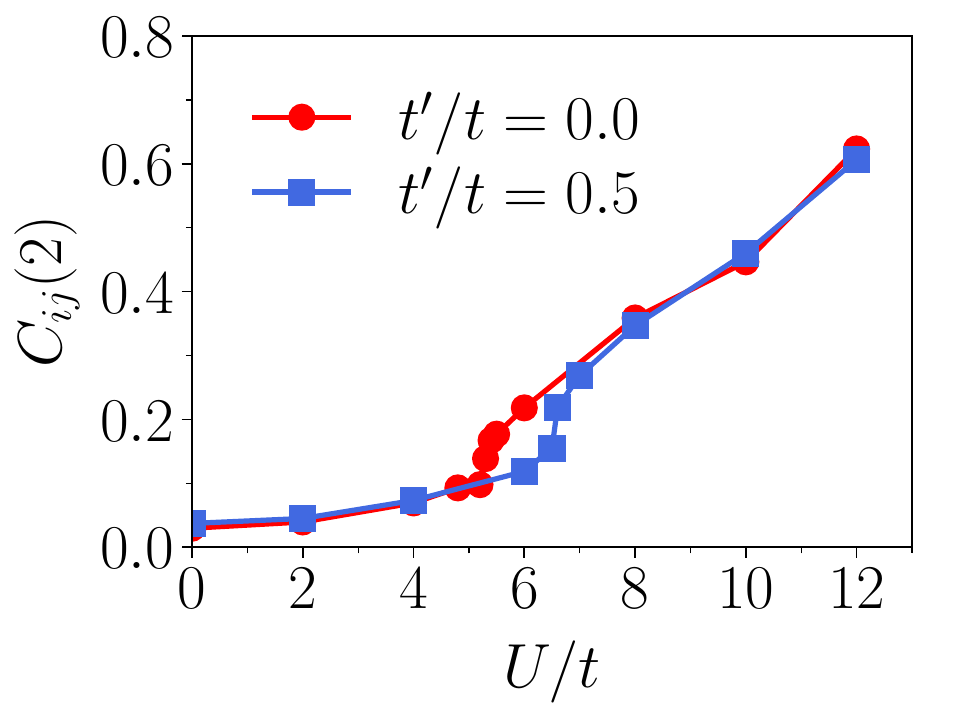}}
\caption{Doublon-holon correlations at (a) nearest-neighbor and (b) next-nearest-neighbor
distances as a function of $U/t$. Results are shown for two values of $t'/t$.}
\label{fig:dh_corr}
\end{figure}

\begin{figure}[!htb]
\centering
\includegraphics[width=0.50\columnwidth]{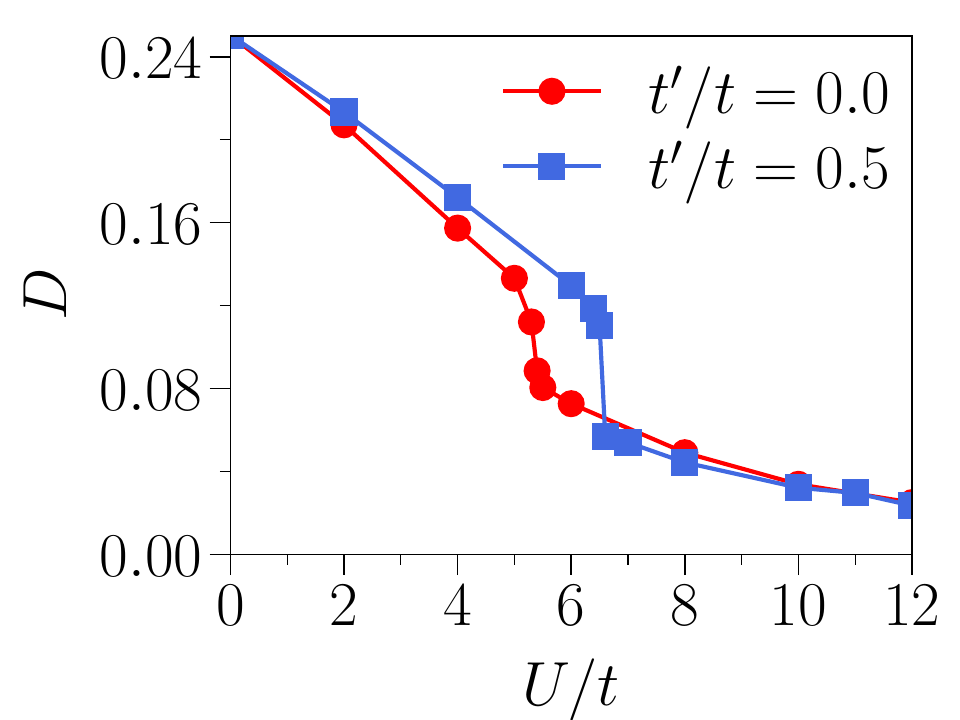}
\caption{Double occupancy $D$ as a function of $U/t$ at the two values of $t'$ shown.}
\label{fig:dbleoccu}
\end{figure}

\begin{figure}[!htb]
\centering
\includegraphics[width=0.50\columnwidth]{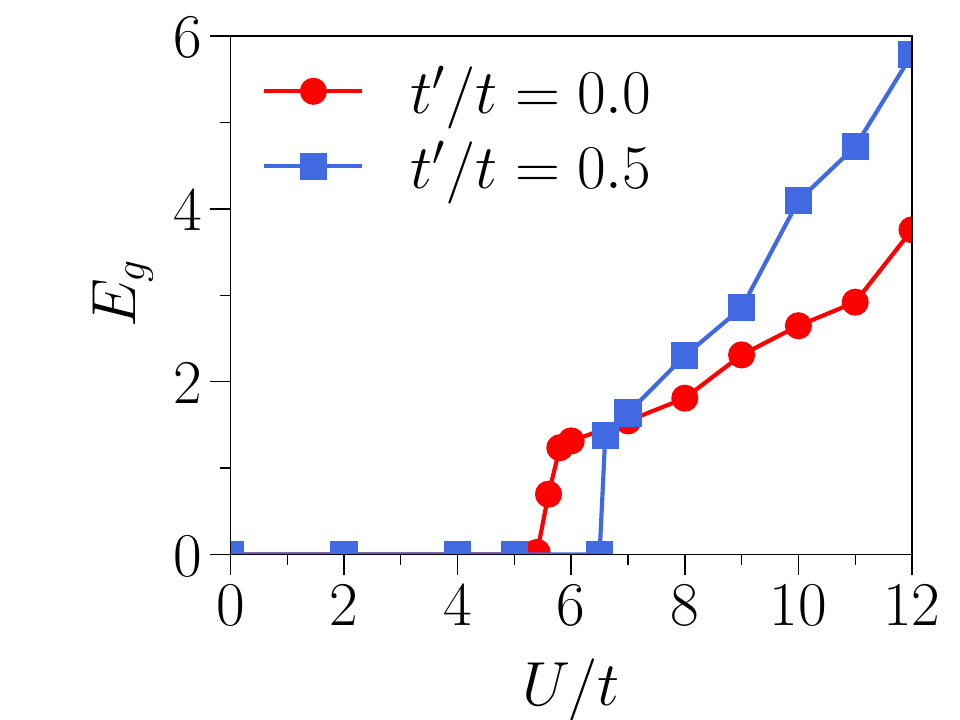}
\caption{Charge gap $E_g$ as a function of $U/t$.}
\label{fig:charge_gap}
\end{figure}

\begin{figure}[!htb]
\centering
\subfigure[\label{fig:nk_a}]{
\includegraphics[width=0.46\columnwidth]{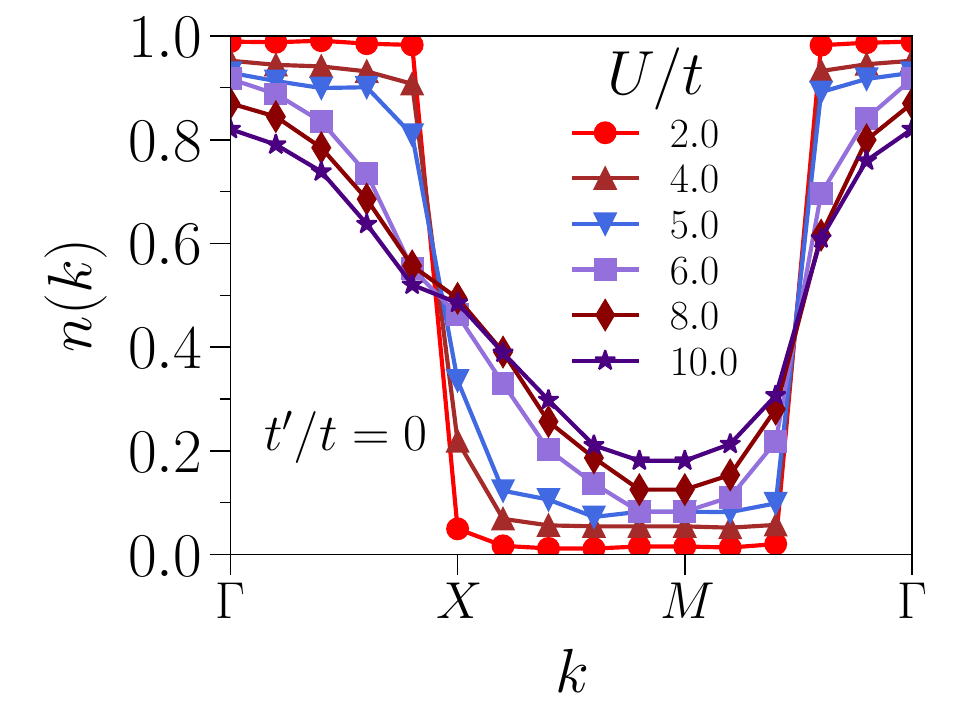}}
\subfigure[\label{fig:nk_b}]{
\includegraphics[width=0.46\columnwidth]{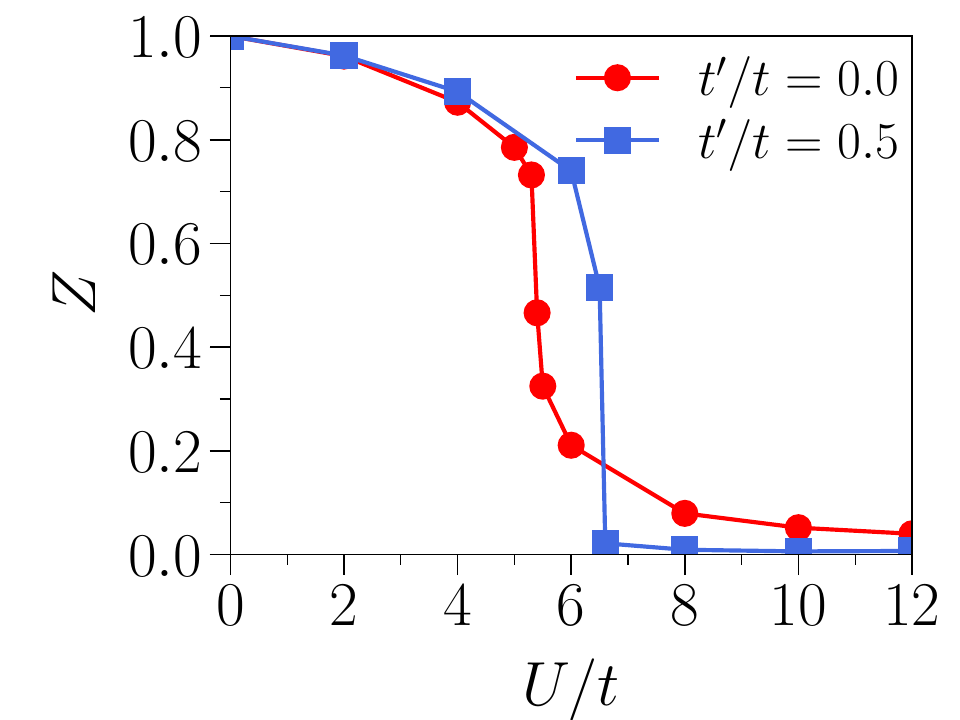}}
\caption{(a) Momentum distribution at various values of $U/t$ along the symmetry path in the first 
Brillouin zone ($t'/t=0$). (b) Quasiparticle weight $Z$ determined from the jump at the Fermi 
surface, as a function of $U/t$, for the two cases of $t'/t$.}
\label{fig:nk}
\end{figure}

\begin{figure}[!htb]
\centering
\subfigure[\label{fig:Sq_a}]{
\includegraphics[width=0.46\columnwidth]{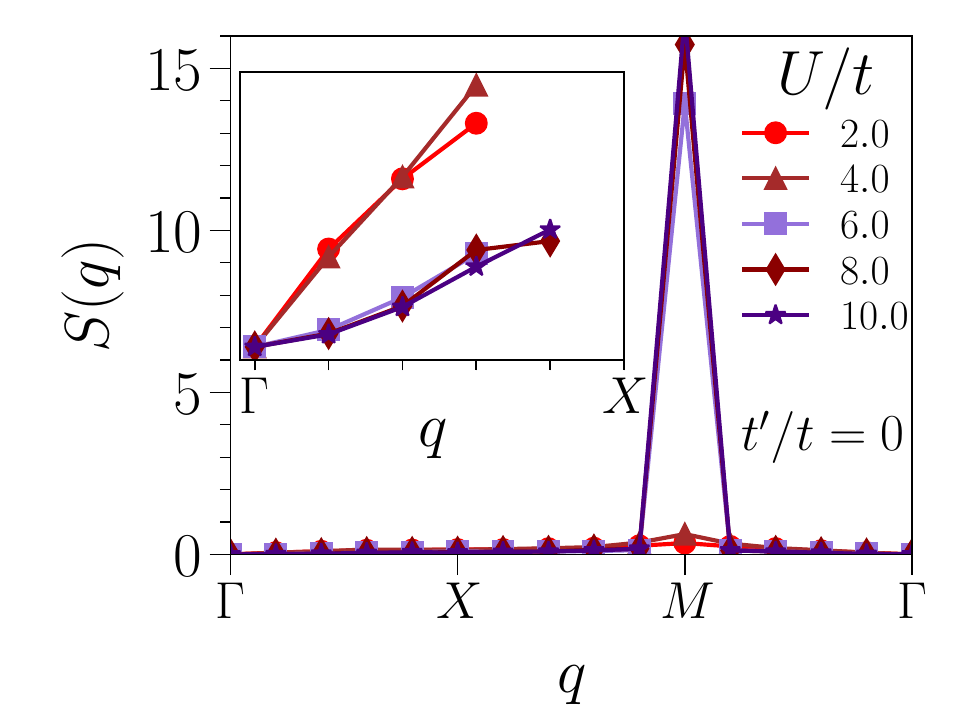}}
\subfigure[\label{fig:Sq_b}]{
\includegraphics[width=0.46\columnwidth]{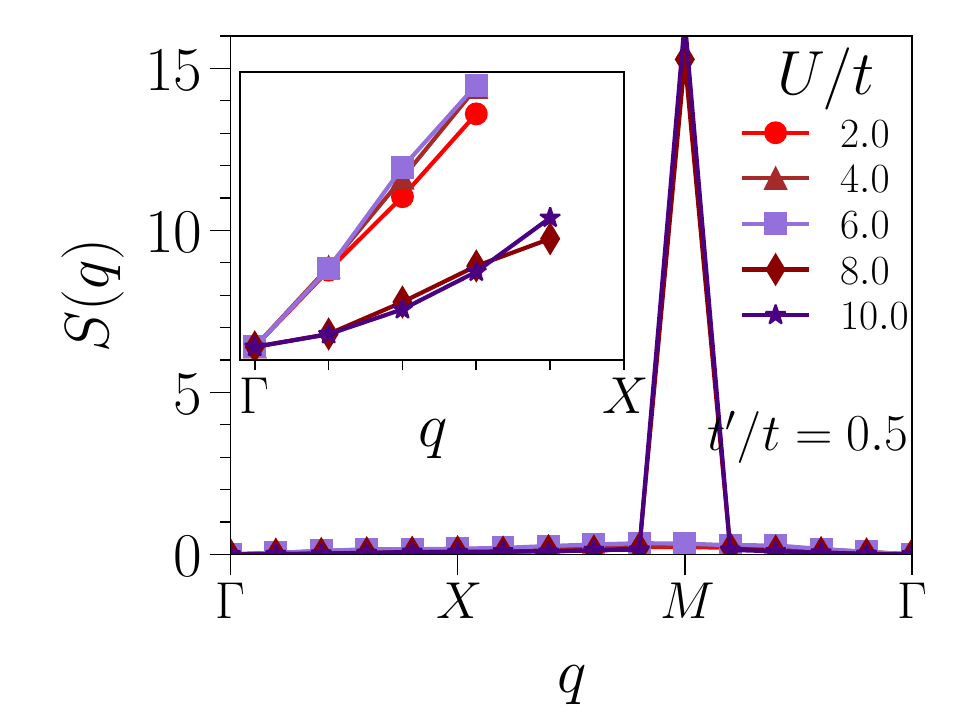}}
\caption{Spin structure factor $S(\vec{q})$ as a function of $\vec{q}$ 
at various values of $U/t$ for the cases (a) $t'/t=0$ and (b) $t'/t=0.5$. The inset in each case
show $S(\vec{q})$ versus $\vec{q}$ for small $|\vec{q}|$. It shows that in the limit 
$|\vec{q}|\rightarrow 0$, $S(\vec{q}) \propto |\vec{q}|$ for $U < U_c$ (gapless), 
while $S(\vec{q}) \propto |\vec{q}|^2$ for $U>U_c$ (spin gap).}
\label{fig:Sq}
\end{figure}

In contrast to the above studies which considered wave functions with different symmetries, 
here we study the ground state within the single variational wave function.
We show that, though no magnetic order is put explicitly into the CRBM+BCS wave function, 
it spontaneously gives rise to long range AF order in the insulating state. This is remarkable 
as this was not observed with any of the other variational WFs considered in the previous studies. 
In order to characterize the ground state, first we compute the doublon-holon (DH) correlation 
function defined as,
\begin{align}
C_{ij}(r) = \frac{\la d_i h_j\ra - \la d_i \ra\la h_j\ra}
{\la d_i \ra\la h_j\ra}
\end{align}
where $d_i=n_{i\up}n_{i\dn}$ and $h_i=(1-n_{i\up})(1-n_{i\dn})$ are the doublon and
holon operator, respectively. The results for the nearest-neighbor (NN) and 
next-nearest-neighbor (NNN) correlations as a function of $U$ are shown in Fig.~\ref{fig:dh_corr}.
It shows that $C_{ij}(r)$ is very close to zero at small $U/t$. As $U/t$ is increased,
it shows a jump at a critical $U_c/t$ and steadily rises after. The NNN values are an order
of magnitude smaller than the NN values, indicating the bindings of doublons and holons 
within short distances from each other. Thus the wave function is able to 
capture the correct physics of doulon-holon bindings in the strong coupling regime in spite of 
it not having any explicit doublon-holon correlation factor. 
Next, we compute the double occupancy, $D$ defined as
\begin{align}
D=\frac{1}{N_{\mathrm{s}}} \sum_{i} n_{i \uparrow} n_{i \downarrow}
\end{align}
It is a crucial quantity that can indicate the presence of Mott transition. The value of $D$
is shown in Fig.~\ref{fig:dbleoccu}.
Starting from a value of $0.25$ at $U/t=0$, $D$ decreases as $U/t$ is increased. It shows a sudden 
drop at a critical $U_c/t$ indicating the onset of Mott transition. The jump in the value
of $D$ is clearly present at both the values of $t'$, though it is sharper in the frustrated case. 
It may be mentioned that, even in the long ranged backflow correlated WF used in 
previous studies\cite{Tocchio_PhysRevB.83.195138}, the $D$ at Mott transition 
at $t'/t=0$ shows only a kink not a jump. We find that the values of critical 
interaction are $U_c/t\sim 5.4$ for $t'/t=0$ and $U_c/t\sim 6.6$ for $t'/t=0.5$. 
These values are in roughly in good agreement 
with previous results\cite{YokoyamaJPSJ2006,Tocchio_PhysRevB.78.041101}. The occurence
of Mott transition can be confirmed directly by looking at the charge gap which can be 
estimated from the knowledge of the ground state WF itself. Given the ground state $\ket{\Psi}$, 
an excited state with momentum $\vec{q}$ is given by $n_{\vec{q}}\ket{\Psi}$, and the
charge gap in limit $q\rightarrow 0$ for square lattice can be shown to 
be\cite{Tocchio_PhysRevB.83.195138}, 
\begin{align}
E_g = -\frac{1}{4}\left(\lim_{q\rightarrow 0} \frac{|\vec{q}|^2}{N(\vec{q})}\right)
\left({\cal K}_1 + 2{\cal K}_2\right)
\end{align}
where $N(\vec{q}) = \la n_{\vec{q}} n_{-\vec{q}} \ra$ is the charge structure factor.  
${\cal K}_1$ and ${\cal K}_2$ are the NN and the NNN kinetic energy per site, respectively.
The results for $E_g$ are shown in Fig.~\ref{fig:charge_gap}.
It confirms that the state below $U_c/t$ is metallic with no charge gap, while above
$U_c/t$ it is insulating with a finite charge gap.
Next we look at the momentum distribution function, 
$n(\vec{k}) = \la c^\dag_{k\sigma}c_{k\sigma}\ra$. The $n(\vec{k})$ values calculated as
a function of $\vec{k}$ along the symmetry path $\Gamma(0,0)$-$X(\pi,0)$-$M(\pi,\pi)$-$\Gamma(0,0)$ 
for different values of $U/t$ for the case $t'/t=0$ are shown in Fig.~\ref{fig:nk_a}.
In the metallic state, $n(\kv)$ has a discontinuity at the Fermi surface, $k=k_F$ in the nodal
$\Gamma-M$ direction. The magnitude of the jump gives the quasiparticle weight $Z$ which roughly
corresponds to the inverse effective mass of the quasiparticles\cite{Paramekanti_PhysRevB.70.054504,
YokoyamaJPSJ2006}. We plot the values of $Z$ so determined as a function of $U/t$ in 
Fig.~\ref{fig:nk_a}. For both the cases of $t'/t$, $Z$ decreases with increasing $U/t$ and show a sudden drop
at the respective critical interaction $U_c/t$. Interestingly, $Z$ does not vanish completely
in the Mott state for the unfrustrated case at $t'/t=0$, though it becomes very small. Thus it
suggests that the Mott transition in this case takes place via vanishing of spectral weight at the 
Fermi level rather than via divergence of effective mass. In contrast, $Z$ vanishes completely 
in the Mott state for $t'/t=0.5$ case suggesting divergence of the effective mass in this case. 
For both the cases, the transition is found to be first-order nature as evidenced from 
discontinuities of various quantities at the critical point.
Finally, we look at the magnetic correlations in the CRBM wave function. We calculate
the spin structure factor $S(\vec{q})$ defined as, 
\begin{align}
S(\vec{q}) = \frac{1}{L^2} \sum_{i,j} e^{i\vec{q}\cdot(\rv_i-\rv_j)}\la S^z_i S^z_j \ra
\end{align}
where $S^z_i=(n_{i\up}-n_{i\dn})$ is the $z$-component of the spin operator at site $i$. The results
for $S(\vec{q})$ calculated at various values of $U/t$ are shown in Fig.~\ref{fig:Sq}.
As the figure shows, $S(\vec{q})$ is very small at all $\vec{q}$ for $U<U_c$ 
indicating very weak magnetic correlations in the metallic state. For $U>U_c$, $S(\vec{q})$ is
very sharply peaked at $\vec{q}=(\pi,\pi)$ which indicates onset on long range AF correlations
as soon as the system enters the Mott state. The non-zero values of $t'$ considered
here does not seem have any impact on the AF correlations. In fact, we find that the
sublattice magnetization $m=|S^z_i|$ which is close zero in the metallic phase, jumps 
to around $0.8$ at transition, a value close to the saturation limit. 
If we compare with results in Ref.~\onlinecite{YokoyamaJPSJ2006}, the AF correlations in
the insulating state in the GW+DH wave function used in this study is much weaker compared
to what is found here. The insets in the figures show 
$S(\vec{q})$ for a small range of $|\vec{q}|$. Clearly in the limit $|\vec{q}|\rightarrow 0$, 
the $S(\vec{q})\propto |\vec{q}|$ in the metallic phase suggesting absence of spin gap in
this phase. On the other hand, $S(\vec{q}) \propto |\vec{q}|^2$ for $U>U_c$ suggesting
that the Mott state is also spin gapped. 

\section{Conclusion}
\label{sec:conclusion}
In summary, we have studied the ground state phase of the half-filled Hubbard model on a square lattice
using a variational wave function which is constructed by applying a convolutional restricted 
Boltzmann machine (CRBM) neural network as correlator to a mean-field BCS wave function. 
The number of variational parameters in the wave function does not automatically grow with
the lattice size and can be tuned. The wave function is also highly accurate, especially in the 
strong coupling limit where it yields variational energies lower than those of the best known
Jastrow variational WFs for the model. 
The picture of Mott metal-insulator transition in the 
model as described by the CRBM wave function is roughly similar to that obtained by using other 
variational wave functions\cite{YokoyamaJPSJ2006,Tocchio_PhysRevB.78.041101,Tocchio_PhysRevB.83.195138},
with some interesting differences. 
Regarding the shortcomings of the CRBM wave function, the results for the energies shows that
the it does not necessarily perform better in the weak coupling limit.
Other more accurate methods strongly suggest that for the unfrustrated model with $t'/t=0$,
there exist short range AF fluctuations even in the weak coupling limit and 
the ground state in this case is insulating at all $U_c/t>0$\cite{Toschi_PhysRevB.91.125109,Gull_PhysRevX.5.041041,SiddharthaLal_NJP2020}.
This physics is not captured correctly in the CRBM wave function. This of course 
can be remedied readily by putting an AF order manually into the mean-field part of the
wave function, though it would be much desirable to have the correlations generated 
spontaneously in the same manner as in the Mott insulating state. It is also interesting to
study the wave function for a range of $t'/t$ values in order to obtain a full phase diagram
as a function of $U/t$ and $t'/t$. We leave it for a future study. 

\section*{Acknowledgement} 
The authors thank the Science and Engineering Research Board, DST, Govt of
India for financial support under the Core grant (No: CRG/2021/005792). 
Also acknowledge CHPC, IISER Thiruvananthapuram for computational facilities. 

\bibliography{references}

\end{document}